\documentclass [twocolumn]{aastex701} 

\newcommand\degree{\degr}
\newcommand\degrees\degree

\newcommand\jwst{\em JWST}

\newcommand\eureka{\texttt{Eureka!}\xspace}

\DeclareSymbolFont{UPM}{U}{eur}{m}{n}
\DeclareMathSymbol{\umu}{0}{UPM}{"16}
\let\oldumu=\umu
\renewcommand\umu{\ifmmode\oldumu\else\math{\oldumu}\fi}

\newcommand\microns \micron

\let\oldsim=\sim
\renewcommand\sim{\ifmmode\oldsim\else\math{\oldsim}\fi}
\let\oldpm=\pm
\renewcommand\pm{\ifmmode\oldpm\else\math{\oldpm}\fi}
\newcommand\by{\ifmmode\times\else\math{\times}\fi}

\newbox{\wdbox}
\renewcommand\c{\setbox\wdbox=\hbox{,}\hspace{\wd\wdbox}}
\renewcommand\i{\setbox\wdbox=\hbox{i}\hspace{\wd\wdbox}}
\newcommand\n{\hspace{0.5em}}

\newcount\timect
\newcount\hourct
\newcount\minct
\newcommand\now{\timect=\time \divide\timect by 60
         \hourct=\timect \multiply\hourct by 60
         \minct=\time \advance\minct by -\hourct
         \number\timect:\ifnum \minct < 10 0\fi\number\minct}

\catcode`@=11

\newcommand\comment[1]{}
\newcommand\commenton{\catcode`\%=14}
\newcommand\commentoff{\catcode`\%=12}

\renewcommand\math[1]{$#1$}
\newcommand\mathshifton{\catcode`\$=3}
\newcommand\mathshiftoff{\catcode`\$=12}

\comment{the backslash is necessary}

\comment{alignment tab}

\let\atab=&
\newcommand\atabon{\catcode`\&=4}
\newcommand\ataboff{\catcode`\&=12}

\let\oldmsp=\sp
\let\oldmsb=\sb
\def\sp#1{\ifmmode
           \oldmsp{#1}%
         \else\strut\raise.85ex\hbox{\scriptsize #1}\fi}
\def\sb#1{\ifmmode
           \oldmsb{#1}%
         \else\strut\raise-.54ex\hbox{\scriptsize #1}\fi}
\newbox\@sp
\newbox\@sb
\def\sbp#1#2{\ifmmode%
           \oldmsb{#1}\oldmsp{#2}%
         \else
           \setbox\@sb=\hbox{\sb{#1}}%
           \setbox\@sp=\hbox{\sp{#2}}%
           \rlap{\copy\@sb}\copy\@sp
           \ifdim \wd\@sb >\wd\@sp
             \hskip -\wd\@sp \hskip \wd\@sb
           \fi
        \fi}
\def\msp#1{\ifmmode
           \oldmsp{#1}
         \else \math{\oldmsp{#1}}\fi}
\def\msb#1{\ifmmode
           \oldmsb{#1}
         \else \math{\oldmsb{#1}}\fi}
\def\supon{\catcode`\^=7}
\def\supoff{\catcode`\^=12}
\def\subon{\catcode`\_=8}
\def\suboff{\catcode`\_=12}
\def\supsubon{\supon \subon}
\def\supsuboff{\supoff \suboff}

\newcommand\actcharon{\catcode`\~=13}
\newcommand\actcharoff{\catcode`\~=12}

\newcommand\paramon{\catcode`\#=6}
\newcommand\paramoff{\catcode`\#=12}

\comment{And now to turn us totally on and off...}

\newcommand\reservedcharson{\commenton \mathshifton \atabon \supsubon \actcharon
	\paramon}

\newcommand\reservedcharsoff{\commentoff \mathshiftoff \ataboff
	\supsuboff \actcharoff \paramoff}

\catcode`@=12
\reservedcharsoff

\reservedcharson

\comment{ Must have ONLY ONE of these... trust these macros, they work

}

\newcommand{\squishlist}{
 \begin{list}{$\bullet$}
  { \setlength{\itemsep}{0pt}
     \setlength{\parsep}{0pt}
     \setlength{\topsep}{0pt}
     \setlength{\partopsep}{0pt}
     \setlength{\leftmargin}{2.0em}
     \setlength{\labelwidth}{1.5em}
     \setlength{\labelsep}{0.5em} } }

\newcommand{\squishlisttwo}{
 \begin{list}{$\bullet$}
  { \setlength{\itemsep}{1pt}
     \setlength{\parsep}{3pt}
     \setlength{\topsep}{3pt}
     \setlength{\partopsep}{0pt}
     \setlength{\leftmargin}{2.0em}
     \setlength{\labelwidth}{1.5em}
     \setlength{\labelsep}{0.5em} } }

\newcommand{\squishend}{
  \end{list}  }

\reservedcharson
\actcharon

\newcommand{\figsetcapnum}{}
\newcommand{\figsetcaptitle}{}

\renewcommand{\figsetgrpstart}{\begin{figure*}[!h]\renewcommand{\figurename}{Fig.}\renewcommand{\thefigure}{Set \figsetcapnum~$-$ \figsetcaptitle}
}
\renewcommand{\figsetgrpend}{\end{figure*}}
\usepackage{amssymb, amsmath, mathtools}
\usepackage{enumerate}
\usepackage{graphicx}

\usepackage{latexsym}
\usepackage[T1]{fontenc}
\usepackage{tabularx}
\usepackage{multirow}
\usepackage{colortbl}
\usepackage[version=4]{mhchem} 
\usepackage{xspace}
\usepackage{color}
\usepackage{graphicx}
\usepackage{hyperref} 
\usepackage[nameinlink]{cleveref}

\graphicspath{{./}{figs/}}
\makeatletter
\def\input@path{{./}{tables/}}
\makeatother

\shorttitle{Evidence for LP 890-9d via Transit Timing Variations}
\shortauthors{Stevenson et al.}

\begin{document}

\title{Evidence for LP 890-9d via Transit Timing Variations}

\correspondingauthor{Kevin B. Stevenson}
\email{Kevin.Stevenson@jhuapl.edu}

\newcommand{\APL}{Johns Hopkins Applied Physics Laboratory, 11100 Johns Hopkins Rd, Laurel, MD 20723, USA}
\newcommand{\Goddard}{NASA Goddard Space Flight Center, 8800 Greenbelt Road, Greenbelt, MD 20771, USA}
\newcommand{\CHAMPs}{Consortium on Habitability and Atmospheres of M-dwarf Planets (CHAMPs), Laurel, MD, USA}
\newcommand{\JHUAstro}{Department of Physics and Astronomy, Johns Hopkins University, Baltimore, MD, USA}

\author[0000-0002-7352-7941]{Kevin B. Stevenson}
\affiliation{\APL}
\email{Kevin.Stevenson@jhuapl.edu}

\author[0000-0002-3263-2251]{Guangwei Fu}
\affiliation{\JHUAstro}
\email{guangweifu@gmail.com}

\author[0000-0001-7393-2368]{Kristin S. Sotzen}
\affiliation{\APL}
\email{Kristin.Sotzen@jhuapl.edu}

\author[0000-0002-2739-1465]{E. M. May}
\affiliation{\APL}
\email{Erin.May@jhuapl.edu}

\author[0000-0002-1664-4105]{Larissa Palethorpe}
\affiliation{School of Physics, University of Bristol, HH Wills Physics Laboratory, Tyndall Avenue, Bristol BS8 1TL, UK}
\email{larissa.palethorpe@bristol.ac.uk}

\author[0000-0002-0746-1980]{Jacob Lustig-Yaeger}
\affiliation{\APL}
\email{Jacob.Lustig-Yaeger@jhuapl.edu}

\author[0000-0002-1570-2203]{Ted M. Johnson}
\affiliation{Nevada Center for Astrophysics, University of Nevada, Las Vegas, 4505 South Maryland Parkway, Las Vegas, NV 89154, USA}
\affiliation{Department of Physics and Astronomy, University of Nevada, Las Vegas, 4505 South Maryland Parkway, Las Vegas, NV 89154, USA}
\email{johnst82@unlv.nevada.edu}

\author[0000-0002-0802-9145]{Eric Agol}
\affiliation{Astronomy Department and Virtual Planetary Laboratory, University of Washington, Seattle, WA 98195, USA}
\email{agol@uw.edu}

\author[orcid=0000-0002-9030-0132]{Katherine A. Bennett}
\affiliation{Department of Earth \& Planetary Sciences, Johns Hopkins University, Baltimore, MD 21218, USA}
\email{kbenne50@jhu.edu} 

\author[0000-0001-5097-9251]{Carlos Gascón}
\affiliation{Space Telescope Science Institute, Baltimore, MD, USA}
\email{cgascon@stsci.edu}

\author[0000-0001-6050-7645]{David K.\ Sing}
\affiliation{Department of Earth \& Planetary Sciences, Johns Hopkins University, Baltimore, MD 21218, USA}
\affiliation{\JHUAstro}
\email{dsing@jhu.edu}

\author[0000-0003-3305-6281]{Jeff A. Valenti}
\affiliation{Space Telescope Science Institute, Baltimore, MD, USA}
\email{valenti@stsci.edu}

\author[0000-0003-4328-3867]{Hannah R. Wakeford}
\affiliation{School of Physics, University of Bristol, HH Wills Physics Laboratory, Tyndall Avenue, Bristol BS8 1TL, UK}
\email{hannah.wakeford@bristol.ac.uk}



\begin{abstract}
LP~890-9, also known as SPECULOOS-2 and TOI-4306, is a nearby late-M dwarf hosting two confirmed transiting rocky exoplanets. We analyze 20 {\jwst}/NIRSpec PRISM transits of LP~890-9b and LP~890-9c obtained as part of GO program 7073 and detect statistically significant transit timing variations (TTVs), with peak-to-peak amplitudes of $\sim 17$~s and $\sim 35$~s, respectively. Using analytic linear TTV theory, we find that the known two-planet configuration cannot reproduce the measured TTV amplitudes or super-period, whereas three-planet models provide substantially better fits. The best-fit configuration places the candidate third planet, LP~890-9d, between planets b and c, with $P_d \sim 4.4$~days; however, the current data do not uniquely determine its orbital architecture and periods spanning 4.0--6.9~days remain plausible. \texttt{TESS} is insensitive to transits of LP~890-9d and we find no evidence for the candidate in {\jwst} observations, although the phase coverage (ranging from $\sim 50\%$ to $\sim 80\%$) depends strongly on the candidate orbital period. Additional high-precision transit observations of LP~890-9b and LP~890-9c are needed to refine their TTV solutions and further constrain the orbital properties of the third planet.
\end{abstract}

\keywords{
\uat{Exoplanet systems}{484};
\uat{Exoplanet dynamics}{490};
\uat{Exoplanets}{498};
\uat{Habitable planets}{695};
\uat{M dwarf stars}{982};
\uat{Transit timing variation method}{1710};
\uat{Transits}{1711};
\uat{Astronomy data analysis}{1858}
}

\section{Introduction}
\label{sec:intro}

\subsection{Transit Timing Variations}

In a Keplerian two-body system, transits occur strictly periodically. The presence of additional bodies, however, introduces gravitational perturbations that can produce measurable deviations from a linear ephemeris. These transit timing variations (TTVs) provide a powerful means of probing the masses and orbital architectures of multi-planet systems, particularly compact systems near mean-motion resonances, where planet-planet interactions are enhanced \citep{Miralda2002, Agol2005}. TTVs are therefore especially useful for characterizing small planets around low-mass stars, for which radial-velocity measurements can be observationally challenging.

TTVs have been used both to measure planet masses and to infer the presence of non-transiting companions. In one of the first demonstrations that TTVs could characterize a tightly packed planetary system, \citet{Lissauer2013} measured the masses of the five inner planets in the Kepler-11 system and placed an upper limit on the mass of the outermost planet. As another example, \citet{JontofHutter2015} used TTVs to measure the mass of the Mars-sized planet Kepler-138b, providing the first density measurement for an exoplanet smaller than Earth. More broadly, \citet{Hadden2017} applied TTV analyses to a large sample of Kepler multi-planet systems, measuring planet masses (and eccentricities) and demonstrating the value of TTVs for population-level studies of small planets. TTVs have also been central to the characterization of the TRAPPIST-1 system. \citet{Agol2021} used hundreds of transits from \textit{Spitzer} and ground-based facilities to determine the masses and densities of all seven Earth-sized planets, showing that the planets have broadly similar, likely rocky compositions with modest volatile inventories.

TTVs can also reveal planets that do not themselves transit. For example, \citet{Ballard2011} identified a non-transiting companion in the Kepler-19 system from the sinusoidal TTVs of Kepler-19b, while also emphasizing that the inferred companion properties were not unique. More recently, \citet{Sun2025} reported the discovery of Kepler-725c, a non-transiting super-Earth inferred from TTVs, and \citet{Bonfanti2026} announced the discovery of TOI-5624f from the TTVs of TOI-5624e, supported by radial-velocity observations.

Despite their power, TTV inversions are often highly degenerate. A given timing signal can frequently be reproduced by multiple combinations of perturber mass, period, eccentricity, and orbital phase, particularly when the observing baseline is short or when conjunction-scale timing structure (i.e., ``chopping'') is not detected. In a systematic reassessment of 12 published systems for which non-transiting planets were claimed to be uniquely characterized from timing data alone, \citet{Lammers2026} found that only two provided compellingly unique solutions. Their analysis highlighted the importance of long temporal baselines, accurate transit-time uncertainties, complementary constraints, and the detection of short-timescale chopping signals in establishing robust TTV-based planet characterizations.

\subsection{LP~890-9 System}

LP~890-9, also known as TOI-4306 or SPECULOOS-2, is a nearby late-type M-dwarf star that hosts two known transiting terrestrial-size planets ($\sim1.3 R_{\oplus}$). With an effective temperature of $2850 \pm 75$~K, LP~890-9 is the second-coolest star known to host planets, after TRAPPIST-1. \citet{Delrez2022} announced the discovery of LP~890-9b and LP~890-9c, with orbital periods of $\sim 2.7$~days and $\sim 8.5$~days, respectively. Planet c receives an incident flux comparable to that of Earth ($T_{eq,c} = 269$~K) and lies within the conservative habitable zone, close to its inner edge. Because of the small stellar radius and the favorable atmospheric-characterization prospects of its temperate planet, \citet{Delrez2022} identified LP~890-9c as one of the most favorable habitable-zone terrestrial planets for follow-up with the James Webb Space Telescope ({\jwst}), after the TRAPPIST-1 planets.


The potential habitability and observability of LP~890-9c have motivated several follow-up studies. For example, \citet{GomezBarrientos2023} showed that {\jwst} transmission spectroscopy could distinguish some high-mean-molecular-weight atmospheric scenarios for LP~890-9c, while \citet{Barnes2025} found that its long-term habitability depends sensitively on its volatile inventory and atmospheric history. These studies underscore the broader interest in the LP~890-9 system as a laboratory for temperate terrestrial planets. 

\subsection{Multiple-Transit Observations With {\jwst}}

The precision of {\jwst} enables a new regime of transit timing for small planets orbiting small stars. While the available ground-based photometry of LP~890-9 was not sufficiently precise to detect the few-second TTVs expected from the known two-planet system \citep{Delrez2022}, the large collecting area and stable pointing of {\jwst} can deliver transit times with order-second precision for favorable targets \citep[e.g.,][]{Moran2023, May2023}. Such high-precision timing will become increasingly common as multiple-transit {\jwst} programs push toward the precision needed for atmospheric reconnaissance of terrestrial planets, particularly those with high-mean-molecular-weight atmospheres. These observations naturally provide repeated, precise transit-time measurements in addition to their spectroscopic constraints. Multiple-transit observations with {\jwst}, therefore, provide an opportunity to search for timing signals that were inaccessible from the ground and to test whether the known planets are dynamically perturbed by an additional companion.

In this work, we use precise {\jwst} transit times of LP~890-9b and c to search for evidence of an additional planet in the system. We compare a two-planet model with two representative three-planet configurations, one in which the candidate planet lies between planets b and c and another in which it lies exterior to planet c. We assess the resulting orbital solutions, planet masses, possible transit depths, and the remaining degeneracies that can be tested with future observations.

\section{Analysis} 
\label{sec:analysis}

\subsection{JWST Data Reduction and Fits} 
\label{sec:analysis:reduction} 

{\jwst} observed 20 transits of LP~890-9b and c from 12 August, 2025 to 13 February, 2026 as part of GO program 7073 (PIs: Lustig-Yaeger \& Stevenson). Each NIRSpec/PRISM observation used five groups per integration and lasted 3.85 hours. We reduced and analyzed the time series data using two independent pipelines: \texttt{Tswift} and {\eureka} \citep{Fu2022,Bell2022}.  The subsections below provide a high-level summary of our reduction steps for each pipeline.

\subsubsection{\texttt{Tswift} Reduction} 
\label{sec:analysis:tswift}

We retrieved the uncalibrated data products from MAST and processed them with the default \texttt{jwst} Stage 1 pipeline. Between the dark-subtraction and ramp-fitting steps, we applied a group-level $1/f$ background subtraction using the unilluminated rows at the top and bottom of the detector as the reference. We read the mid-integration times in BJD$_{\rm TDB}$ directly from the \texttt{INT\_TIMES} extension. We identified time-variable outlier pixels with a per-pixel time-series median-absolute-deviation clip at $5{\sigma}$ and replaced flagged samples with \texttt{NaN}s.

We constructed each visit's white-light curve by summing the cleaned 2D cube over detector columns 54--440, corresponding to 0.6--5.3\,{\microns}, using a three-row spatial aperture centered on the spectral trace. After normalizing each light curve by the median raw counts of the pre-ingress integrations, we masked the first few hundred integrations, which contain the strongest nonlinear ramp. We then fit all visits jointly using the \texttt{batman} transit model \citep{Kreidberg2015} with quadratic limb darkening. In the fit, the geometric parameters $R_p/R_\star$, $a/R_\star$, $i$, and $u_2$ were shared across visits, while a linear baseline slope, multiplicative constant, and transit mid-time $t_c$ varied independently for each visit. For the limb-darkening treatment, we fixed $u_1$ using PHOENIX models computed with \texttt{exotic-ld} \citep{Grant2024}, adopting $\mu_{\rm min}=0.2$, $T_{\rm eff}=2850$\,K, $\log g=5.126$, and $[{\rm M}/{\rm H}]=-0.028$. The joint fit used \texttt{emcee} \citep{emcee}, with 64 walkers, a 2,000-step burn-in, and 10,000 production steps.

Because the residuals show time-correlated structure, the formal MCMC posteriors likely underestimate the transit-time uncertainties. We therefore estimated per-visit timing uncertainties using a prayer-bead, or circular residual-permutation, bootstrap \citep{Bouchy2005}. For each visit, we cyclically shifted the best-fit residuals by a random number of integrations, added the shifted residuals back to the best-fit model, and refit only the transit mid-time. We repeated this procedure 1,000 times and adopted the standard deviation of the resulting $t_c$ distribution as our uncertainty. All transit-time posterior distributions are better described by Gaussian distributions than by Student’s-$t$ distributions.
For the eight visits of LP~890-9b, the resulting uncertainties are 1.6--2.4\,s and, for the twelve visits of LP~890-9c, they are 1.7--3.9\,s. This is approximately twice the formal MCMC posterior uncertainties.

\begin{table*}[t]
\centering
\caption{Transit times and TTVs for LP 890-9b and LP 890-9c from \texttt{Tswift} and {\eureka} analyses. MJD = BJD - 2,460,900.}
\label{tab:transit_times}
\begin{tabular}{rrrrrrr}
\hline
$t0_{\,\mathrm{Tswift}}$ & $\mathrm{TTV}_{\mathrm{Tswift}}$ & $\sigma_{\mathrm{Tswift}}$ & $t0_{\mathrm{Eureka!}}$ & $\mathrm{TTV}_{\mathrm{Eureka!}}$ & $\sigma_{\mathrm{Eureka!}}$ & $\Delta\mathrm{TTV}$ \\
(MJD) & (sec) & (sec) & (MJD) & (sec) & (sec) & (sec) \\
\hline
LP 890-9b \\
\hline
0.133483  & -4.2 & 2.1 & 0.133481  & -4.4 & 2.7 & 0.2 \\
5.593435  &  8.7 & 1.6 & 5.593418  &  7.2 & 2.1 & 1.5 \\
11.053098 & -3.4 & 1.9 & 11.053102 & -3.0 & 2.2 & -0.3 \\
19.242772 & -6.0 & 1.6 & 19.242769 & -6.2 & 2.2 & 0.3 \\
30.162588 & 12.2 & 1.7 & 30.162575 & 11.1 & 2.2 & 1.1 \\
32.892342 & -0.5 & 2.1 & 32.892352 &  0.4 & 2.4 & -0.9 \\
35.622241 & -0.7 & 2.4 & 35.622225 & -2.1 & 2.3 & 1.4 \\
43.811914 & -3.4 & 1.9 & 43.811922 & -2.7 & 2.2 & -0.7 \\
\hline
LP 890-9c \\
\hline
7.720204   &  -5.3 & 2.7 &   7.720242 &  -2.0 & 2.5 & -3.3 \\
33.092413  &  -5.9 & 3.0 &  33.092376 &  -9.1 & 2.2 & 3.2 \\
50.007090  & -17.5 & 3.9 &  50.007118 & -15.0 & 2.2 & -2.5 \\
66.922200  &   8.3 & 2.6 &  66.922174 &   6.1 & 1.8 & 2.3 \\
75.379655  &  12.6 & 2.6 &  75.379637 &  11.0 & 2.5 & 1.6 \\
83.837149  &  20.3 & 2.9 &  83.837147 &  20.1 & 3.0 & 0.1 \\
100.751786 &   5.2 & 3.3 & 100.751769 &   3.7 & 1.6 & 1.5 \\
134.581165 & -15.8 & 2.5 & 134.581190 & -13.6 & 1.7 & -2.2 \\
143.038707 &  -4.0 & 3.0 & 143.038708 &  -3.9 & 2.2 & -0.1 \\
159.953528 &  -3.1 & 3.4 & 159.953473 &  -7.9 & 2.6 & 4.8 \\
168.411078 &   9.4 & 1.6 & 168.411054 &   7.3 & 2.5 & 2.1 \\
185.325844 &   5.5 & 3.1 & 185.325872 &   7.9 & 2.0 & -2.4 \\
\hline
\end{tabular}
\end{table*}

\subsubsection{Eureka! Reduction} 
\label{sec:analysis:eureka}

We used the {\eureka} pipeline \citep[v1.3;][]{Bell2022} to reduce the {\jwst} NIRSpec/PRISM data and generate white-light curves. For Stages 1 and 2, we converted ramps to slopes and applied standard calibration corrections. We used the \texttt{jwst} pipeline version 1.20.2 with CRDS reference context (i.e., {\em pmap}) 1364. For planet b we skipped the jump detection step and used the top and bottom 6 pixels for the group-level background subtraction to remove the 1/f noise. For planet c we used the {\eureka} optimizer \citep{Ashtari2025} to determine the best jump detection threshold (4$\sigma$) and applied group-level background subtraction using the lower and upper 11 pixels of the 32-pixel tall subarray.

For Stages 3 and 4, we generated a time series of 1D spectra and combined them to form white light curves.
For both planets, we performed sweeps of 14 parameters to determine their optimal values, as determined by minimizing the median absolute difference (MAD) of the white-light curve. For planet b, we performed this independently for each visit to test for consistency of parameters across multiple visits of the same system. For planet c, we optimized only the first visit and adopted the same parameters for all remaining visits.  The optimal aperture size is 7 pixels (full-width). 

For both planets, we fit the band-integrated (0.7 - 5.3 {\microns}) light curves using either a \texttt{batman} or \texttt{fleck} transit model \citep{Kreidberg2015, fleck} and a linear trend in time.  There is no discernible stellar activity and we did not flag any spot crossing events in either the LP~890-9b or LP~890-9c light curves. We fit the eight white-light curves of planet b simultaneously and, separately, the 12 white-light curves of planet c, allowing the transit time and planet-to-star radius ratio to vary independently for each visit. For a given planet, we fit a common $a/R_S$ and inclination over all visits. For planets b and c, we fix the limb darkening values ($u_1=0.111$, $u_2=0.336$) using ExoTiC-LD with a 2850~K PHOENIX model \citep{Grant2024}.

At the $\sim$5 minute ingress/egress timescale, the time-averaged residuals have RMS values of 52--75~ppm for LP~890-9b and 32--65~ppm for LP~890-9c, compared with Gaussian white-noise expectations of 27--30~ppm. This corresponds to excess correlated noise factors of 1.9--2.5 for LP~890-9b and 1.1--2.2 for LP~890-9c. To account for correlated noise on ingress/egress timescales, we inflated the transit-time uncertainty for each visit by the corresponding excess-noise factor.

To test the robustness of our transit times, we also fit Visits 02 and 05 of planet b with a Gaussian Process (GP). These visits exhibit the largest TTVs and are therefore particularly important for distinguishing the \textit{bdc} and \textit{bcd} configurations. We followed the same fitting procedure described above, but include a GP with a Mat\'ern 3/2 kernel as a function of time, as implemented using the \texttt{celerite} package \citep{celerite}. We find the transit time is consistent to +0.7 seconds (0.3$\sigma$) for Visit 02 and -1.5 seconds (0.6$\sigma$) for Visit 05. The GP-derived uncertainties are also smaller than the red-noise-inflated uncertainties, so we retain the more conservative timing uncertainties reported in \autoref{tab:transit_times}.


\subsection{Measured Transit Times and TTVs}
\label{sec:analysis:tts}

Both of our reductions achieve excellent agreement in their measured white-light curve transit times (\autoref{tab:transit_times}); however, the corresponding TTV uncertainties differ slightly because the two analyses adopt slightly different methodological choices and use different approaches to account for red noise.  

\begin{figure*}
    \centering
    \includegraphics[width=1\linewidth]{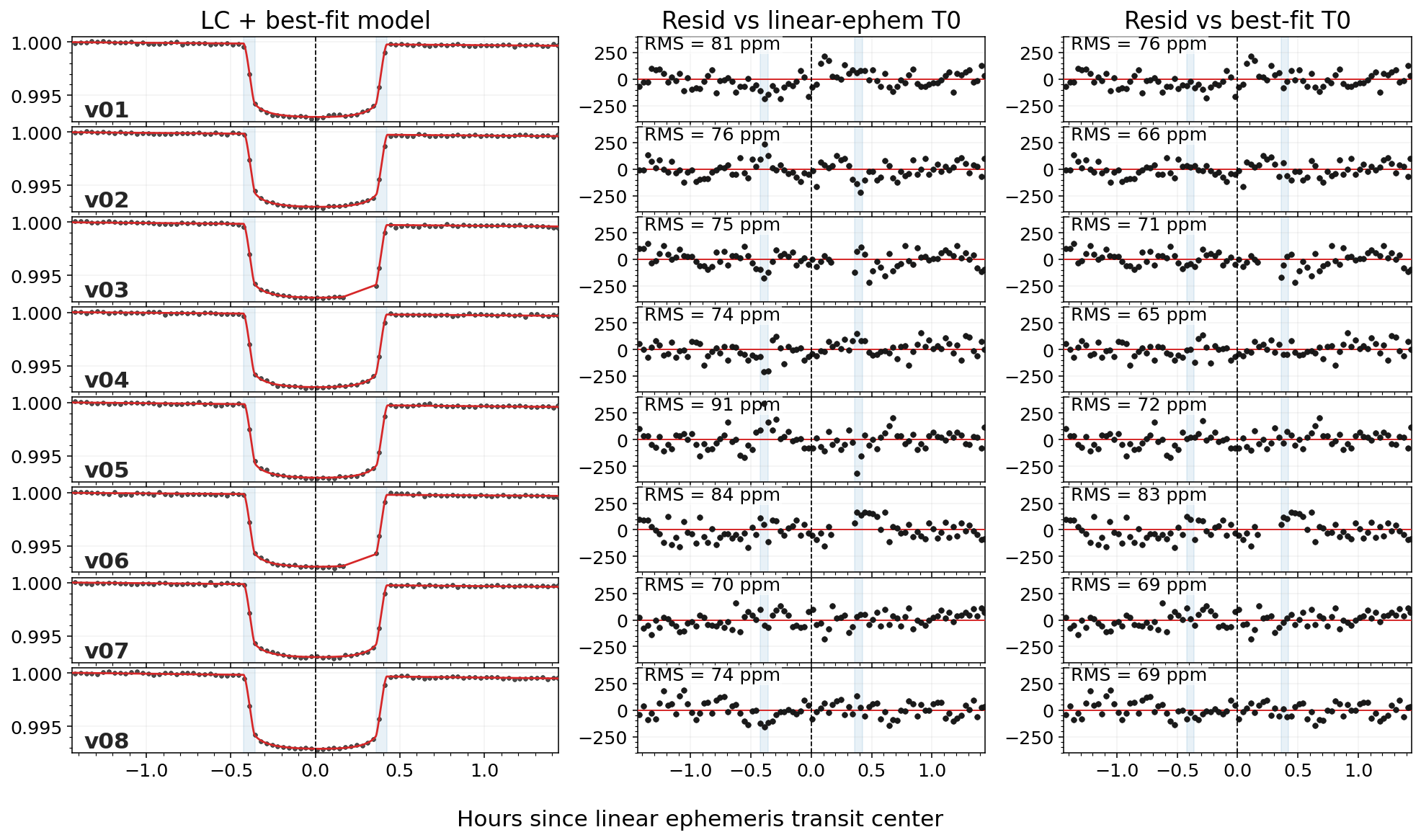}
    \caption{White-light transit fits and residuals (using \texttt{Tswift}) for the eight {\jwst}/NIRSpec PRISM transits of LP~890-9b. Each row corresponds to an individual visit. The left panels show the normalized white-light curves and best-fit transit models; the TTVs are small compared with the transit duration and are therefore not readily apparent by eye. The center panels show residuals after subtracting a model fixed to the linear-ephemeris transit time, while the right panels show residuals after allowing the transit time, $t_0$, to vary. In comparing the fixed-ephemeris and best-fit-$t_0$ residuals, the TTVs are only apparent due to the slight improvement in the residual structure during ingress and egress (marked by vertical blue bands).}
    \label{fig:river_b}
\end{figure*}

\begin{figure*}
    \centering
    \includegraphics[width=1\linewidth]{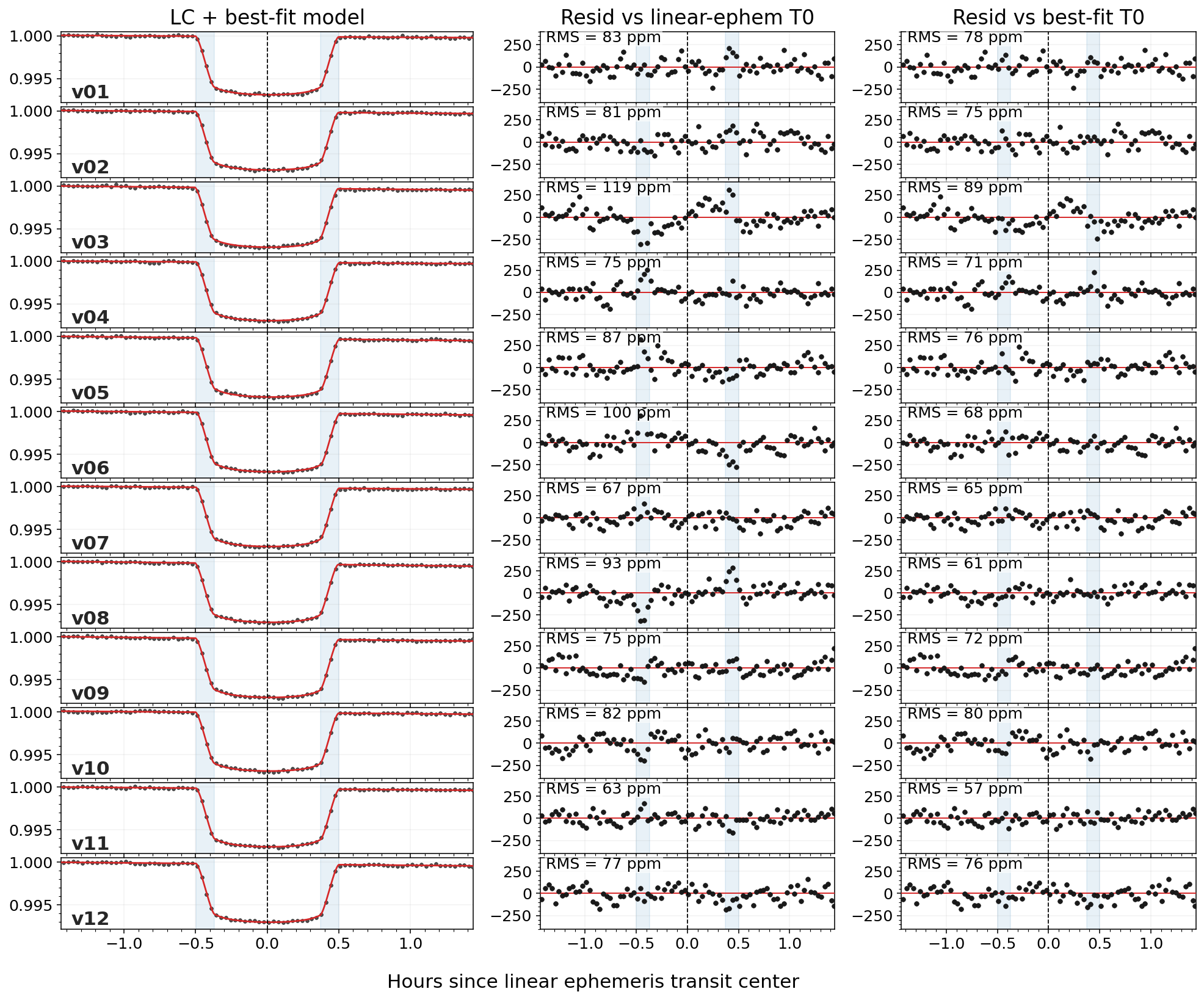}
    \caption{White-light transit fits and residuals (using \texttt{Tswift}) for the 12 {\jwst}/NIRSpec PRISM transits of LP~890-9c. See the caption to \autoref{fig:river_b} for a full description.}
    \label{fig:river_c}
\end{figure*}

\Cref{fig:river_b,fig:river_c} show the \texttt{Tswift} best-fit white-light curves for LP~890-9b and LP~890-9c, respectively, along with residuals computed under two assumptions: a fixed linear ephemeris and an independently fitted transit time for each visit. As an illustrative example, the eighth visit of LP~890-9c (v08) shows negative residuals during ingress and positive residuals during egress when the model is fixed to the linear ephemeris, a pattern indicative of a transit-time offset. Allowing the transit time to vary reduces the residual RMS from 93 to 65~ppm and yields a TTV of $\sim 20$~s. Similar residual patterns are present in other visits to varying degrees, depending on the magnitude of the corresponding TTV.

Using the available ground-based transit photometry, \citet{Delrez2022} searched for, but did not detect, significant TTVs. This non-detection is expected; their typical timing uncertainty is approximately one minute, whereas the predicted maximum TTV amplitudes are only 1.0--1.5~s. The JWST transit times are significantly more precise, with uncertainties comparable to the predicted TTV amplitudes.  As shown in \Cref{fig:ttvs}, the measured transit times show coherent, statistically significant TTVs with amplitudes larger than expected from the nominal two-planet architecture \citet{Delrez2022}.

\subsection{Constraints from the TTV Super-Period}
\label{sec:analysis:degeneracy} 

Transit timing variations provide a powerful means of inferring unseen planetary companions, but the associated inverse problem is often highly degenerate. As discussed previously, \citet{Lammers2026} emphasize that conjunction-scale timing structure (i.e., the chopping signal) is likely necessary, though not sufficient, for uniqueness, and they highlight the importance of long temporal baselines and accurate timing uncertainties. With this in mind, we seek to understand the extent of planet d's orbital period degeneracy.

Using the {\eureka} transit-time measurements for planet c, we first computed TTV residuals with respect to a fixed ephemeris defined by the adopted planet-c period and reference epoch. We then fitted a sinusoidal model to those residuals over a finely sampled grid of trial super-periods, using weighted linear least squares to extract the best-fit amplitude and phase at each trial frequency. The results show a best-fit semi-amplitude of about 12.5 seconds and a super-period near 105.46 days.

\begin{figure}
    \centering
    \includegraphics[width=1\linewidth]{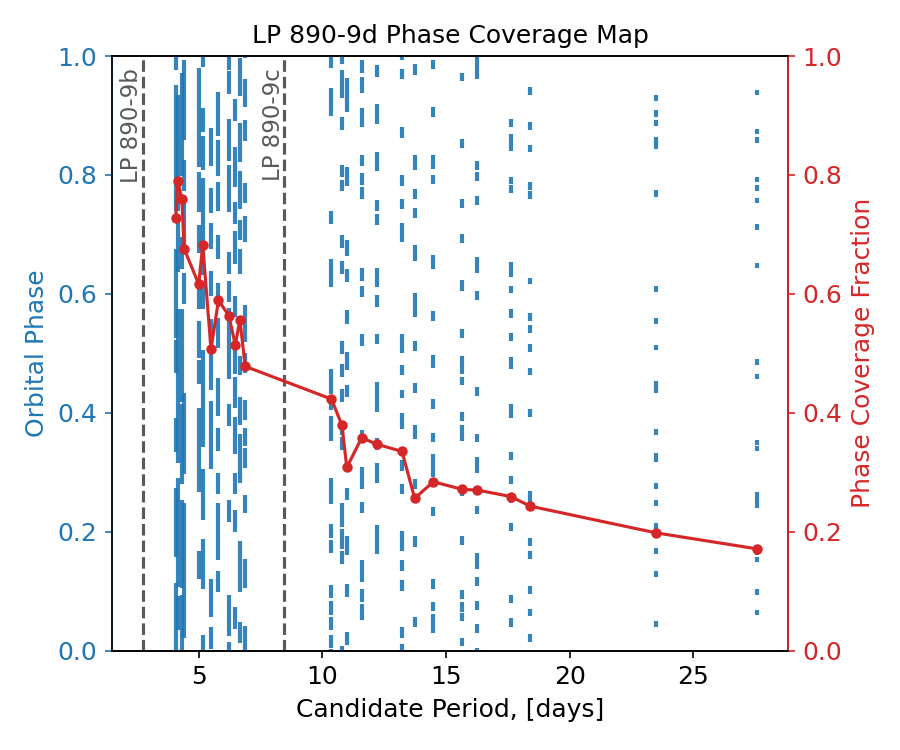}
    \caption{Orbital-phase coverage of the existing {\jwst} observations for the 26 candidate orbital periods of LP~890-9d. Blue vertical segments mark the orbital phases that fall within at least one archived {\jwst} observing window for each candidate period, assuming that LP~890-9d is transiting. The red points and connecting line show the corresponding phase-coverage fraction. Short-period candidates ($P_d = 4-7$~days) have coverage fractions of $\sim 0.5-0.8$, implying that a transit would have had a relatively high probability of occurring during the existing observations. The coverage fraction decreases toward longer periods, leaving larger unobserved phase gaps and making non-detections less constraining for solutions exterior to the orbit of LP~890-9c.}
    \label{fig:phaseCoverage}
\end{figure}

Next, we scanned a dense grid of candidate planet-d periods and low-order resonance indices $j$ and $k$, computing the implied resonant TTV super-period, $P_{TTV}$, for each combination and retaining only those solutions whose super-periods lie close to the best-fit value from the sinusoid. The scan produced a phase-space table of resonance candidates, which was then clustered into families based on contiguous sequences in $P_d/P_c$ for each resonance order. From the filtered phase-space search, we identified 26 distinct families (see \autoref{tab:pd_resonance} in the Appendix).  The 12 candidate solutions interior to planet c range in orbital period from 4.066 -- 6.876 days; the 14 candidate solutions exterior to planet c range from 10.364 -- 27.585 days (see \autoref{fig:phaseCoverage}).
This exercise provided an initial constraint on the number and range of candidate orbital periods for planet d that we then used to search for transits of planet d in the {\jwst} and TESS data 
(\autoref{sec:analysis:jwst} and 
\autoref{sec:analysis:tess}, respectively).

\section{Three-Planet Dynamical Solutions}
\label{sec:3body}

\subsection{Planet Mass Priors} 
\label{sec:analysis:mass} 

We imposed planet mass priors to guide the transit timing fits in \autoref{sec:analysis:ttv2f2f}; however, because the confirmed planets lack direct mass constraints, we estimated their masses using the mass–density–radius relation for small planets from \citet{Luque2022}. We first derived planet radii from the mean transit depths measured in the JWST white-light curves, adopting a stellar radius of $0.1532^{+0.0048}_{-0.0024}\,R{\odot}$ \citep{Delrez2022}. We note that the uncertainties in the planet radii are completely determined by the uncertainty in the stellar radius. We then estimated planet masses using \texttt{Spright} \citep{Spright}, a package that implements a fast Bayesian mass–density–radius relation. To account for asymmetric uncertainties, we provided \texttt{Spright} with asymmetric normal distributions of planet radii and drew 10,000 samples to construct mass distributions for each planet. In \autoref{tab:radii_masses}, we report the median (50th percentile) as the planet mass prior, with the 16th and 84th percentiles defining the corresponding uncertainties. Our transit-time dynamical fitting analysis (\autoref{sec:analysis:ttv2f2f}) requires planet-to-star mass ratios, which we computed using the stellar mass of $0.118 \pm 0.002\,M_{\odot}$ \citep{Delrez2022}.  In this case, the mass ratio uncertainties are dominated by the planet mass uncertainties, thus any improvement in the priors would require a more precise stellar radius. As a final step, we symmetrized the uncertainties when adopting these mass ratios as priors in our subsequent transit-time dynamical fits.

\begin{table}[ht]
\centering
\caption{Median transit depths, radii, masses, and planet-to-star mass ratios for LP 890-9b and LP 890-9c. Uncertainties are quoted as 68\% credible intervals.}
\label{tab:radii_masses}
\begin{tabular}{lcc}
\hline
Measurement & LP 890-9b & LP 890-9c \\
\hline
Transit Depth (ppm) & $6323 \pm 12$ & $6514 \pm 7$ \\
Planet Radius ($R_\oplus$) & $1.328^{+0.042}_{-0.021}$ & $1.348^{+0.042}_{-0.021}$ \\
\hline
Derived Prior & & \\
\hline
Planet Mass Prior ($M_\oplus$) & $2.47^{+0.36}_{-0.29}$ & $2.62^{+0.39}_{-0.33}$ \\
Mass Ratio Prior ($\times 10^{-5}$) & $6.30 \pm 0.84$ & $6.66 \pm 0.92$ \\
\hline
\end{tabular}
\end{table}

\begin{figure*}[t]
    \centering
    \includegraphics[width=\textwidth]{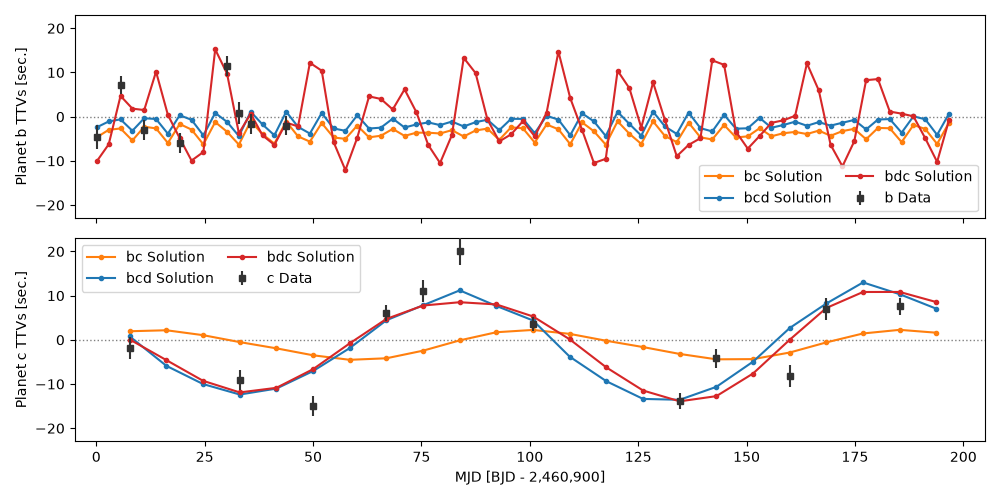}
    \caption{Inferred {\eureka} TTVs for LP~890-9b (top) and LP~890-9c (bottom), shown as black squares, compared with the best-fit (MAP) model solutions, shown as colored points connected by lines. The two-planet {\em bc Solution} cannot adequately reproduce the observed TTVs. Both the {\em bdc} and {\em bcd Solutions} provide more reasonable fits to the planet c TTVs; however, only the {\em bdc Solution} produces amplitudes large enough to match the planet b TTVs. Although none of the solutions perfectly fits the data, the favored orbital configuration places LP~890-9d between planets b and c. We include a similar figure for \texttt{Tswift} in the Appendix (\Cref{fig:ttvs_ttvfast}).
    }
    \label{fig:ttvs}
\end{figure*}

\subsection{Transit-Time Dynamical Fitting with \texttt{TTV2Fast2Furious}} 
\label{sec:analysis:ttv2f2f}

We performed the dynamical inference by fitting the measured transit times directly. This approach allows the linear ephemerides and the dynamical perturbations to be fit simultaneously and avoids neglecting the covariance between the TTV signal and the fitted ephemeris. We generated model transit times and TTVs for the LP~890-9 system using the \texttt{TTV2Fast2Furious} package \citep{ttv2fast2furious}, which implements analytic linear TTV theory. For each set of orbital parameters (periods, transit epochs, masses, eccentricities, and longitudes of periapse), we used \texttt{TTV2Fast2Furious} to construct basis function matrices ($M_{i=1..N}$) that encode the dynamical structure of an N-planet system, including gravitational interactions between all planet pairs. We then used \texttt{TTV2Fast2Furious} to compute the linear model amplitudes ($X_{i=1..N}$) corresponding to the specified orbital configuration. We obtained model transit times for each planet by projecting the full basis matrix onto the amplitude vector ($M_i \cdot X_i$). We then extracted the TTVs by projecting only the columns associated with perturbations beyond the linear ephemeris (i.e., ignoring the first two columns) onto the same amplitude vector  ($M_i[:,2:] \cdot X_i[2:]$). This approach provides an efficient and accurate representation of multi-body gravitational effects on transit times for systems with small eccentricities and planetary masses.
Note that the first two entries of each amplitude vector are set directly to the same trial $t_0$ and $P$ values used to construct the basis, so the linear ephemeris and perturbation basis are evaluated self-consistently. 

We performed a global optimization using differential evolution to obtain initial best-fit parameters. We defined the objective function as the negative log-probability, combining a Gaussian likelihood (based on the agreement between observed and model transit times) with broad, uniform priors on all fitted parameters (masses, transit epochs, periods, $h=e\cos(\omega)$, and $k=e\sin(\omega)$).\footnote{\(e\) is the orbital eccentricity, \(\omega\) is the argument of periapsis, and \(h\) and \(k\) are the components of the eccentricity vector, which fit more smoothly when the eccentricity is small.}
We ran up to 250 iterations of differential evolution, using a fixed random seed for reproducibility, to explore the parameter space and identify the global minimum of the objective function. This approach enabled efficient exploration of our high-dimensional parameter space in which the orbital parameters of planet d were unconstrained.

Following the least-squares optimization, we performed Bayesian inference using affine-invariant ensemble MCMC sampling with \texttt{emcee} \citep{emcee}. We initialized 50 walkers near the best-fit solution, discarded at least the first 80,000 steps as burn-in (due to the lower, \sim3\%, acceptance rates), and then sampled the posterior for 100,000 steps. The sampler employed stretch moves with a scale parameter of 1.2 to improve the proposal acceptance rate \citep{goodman2010,emcee}.
For planets b and c, we adopted normal priors for their masses (as described in \autoref{sec:analysis:mass}), transit epochs, and periods, and uniform priors for the remaining parameters. We estimated posterior distributions from the flattened chains after burn-in, reporting median values and 68\% credible intervals (16th and 84th percentiles) as the associated uncertainties.

\subsubsection{\texttt{TTV2Fast2Furious} Results} 
\label{sec:results}

In \autoref{sec:analysis:degeneracy}, we identified a strong degeneracy in the orbital period of planet d, with 12 and 14 plausible solutions interior and exterior to planet c, respectively. Using our differential-evolution global optimization strategy on the {\eureka} measurements, we converged on a single best-fit solution for each planetary ordering, hereafter referred to as the {\em bdc} and {\em bcd Solutions}. We emphasize, however, that our methodology does not rule out other possible orbital periods for planet d. Therefore, the results presented below should be interpreted as representative solutions, rather than as a definitive characterization of the LP~890-9 system.  
To establish a baseline for comparison, we first consider the {\em bc Solution}, wherein we assume planet d does not exist.

\subsubsection{bc Solution}
\label{sec:results:bc}

Our two-planet, {\em bc Solution} yields a relatively small peak-to-peak amplitude of 5.3~s for planet b and a somewhat larger variation of 6.8~s for planet c. Attempts to fit the latter transit times require a solution with a relatively high mass for planet b and a similarly low mass for planet c (\Cref{tab:orbital_params}). Furthermore, the {\em bc Solution} super-period ($\sim84.6$~days) is inconsistent with the observed super-period of 105.46 days (\autoref{sec:analysis:degeneracy}). Compared with the three-planet solutions, this model provides the poorest fit to the measured transit times.

For a two-planet interacting system, the planets' TTVs are generally expected to be anti-correlated because energy and angular-momentum exchange causes their orbital periods to vary in opposite directions \citep{Agol2005,Holman2005}. The current LP~890-9b timings do not show a clear anti-correlation with the LP~890-9c TTVs, which further disfavors the two-planet interpretation; however, the planet-b baseline spans only 43 days (40\% of the inferred super-period), so the present data cannot definitively establish whether or not such an anti-correlation is present. In the next two subsections, we explore whether an additional perturber in different orbital configurations can reproduce the observed TTVs.

\subsubsection{bdc Solution}
\label{sec:results:bdc}

The differential-evolution least-squares minimizer explored the full range of orbital periods between those of planets b and c. The resulting best-fit period of $\sim 4.40$~days is consistent with one of the super-period solutions identified in \autoref{sec:analysis:degeneracy}, at 4.4053~days. The subsequent \texttt{emcee} fit refined this value to $4.416^{+0.011}_{-0.008}$~days (\autoref{tab:orbital_params}).
Adding a third planet in this configuration produces larger peak-to-peak TTV amplitudes of 27.3 and 24.8~s for planets b and c, respectively. 
These models provide the best fit to the {\jwst} transit times, with $\chi^2/N_{\rm data}=3.9$ (\autoref{fig:ttvs}), primarily because they reproduce the two significantly non-zero TTV measurements of planet b. We note that the two independent reductions yield consistent transit times for these visits, and that the corresponding light curves show no unusual outliers.

The derived masses for b and c are consistent with the adopted priors in \autoref{tab:radii_masses}. This is expected given the absence of LP~890-9d transits.
Assuming that planet d has an Earth-like bulk density, its median mass of $0.24^{+0.09}_{-0.10}\,M_\oplus$ implies a radius of $R_d = 0.62^{+0.07}_{-0.10}\,R_\oplus$. Such a radius would make LP~890-9d between that of Mars and Venus.  The corresponding transit depth, $1350^{+320}_{-400}$~ppm, is substantially smaller than the two known transiting planets and below the standard $7\sigma$ TESS detection threshold.

\subsubsection{bcd Solution}
\label{sec:results:bcd}

Following the same strategy, the least-squares minimizer explored orbital periods beyond that of planet c. Again, the resulting best-fit period of $\sim 18.4$~days is consistent with one of the degenerate solutions identified in \autoref{sec:analysis:degeneracy}, at 18.390~days. The subsequent \texttt{emcee} fit refined this value to $18.53 \pm 0.06$~days (\autoref{tab:orbital_params}).
Adding a third planet exterior to planet c produces a reasonable peak-to-peak TTV amplitude for planet c, 26.6~s, in agreement with the {\em bdc Solution}. However, as shown in \autoref{fig:ttvs}, the planet b TTV amplitudes are nearly identical to those from the {\em bc Solution}.  We conclude that this configuration cannot reproduce the measured TTVs of planet b.

The derived masses for b and c are consistent with those derived in the {\em bdc Solution} and the priors.
Assuming that planet d has an Earth-like bulk density, its mass of $0.15^{+0.10}_{-0.05}\,M_\oplus$ implies a radius of $R_d = 0.53^{+0.10}_{-0.07}\,R_\oplus$. This range is slightly less than, but still overlaps with, the {\em bdc Solution} range.  As such, the corresponding transit depth, $980^{+410}_{-240}$~ppm, remains below the standard $7\sigma$ TESS detection threshold.



\begin{table*}[t]
\centering
\caption{Median-fit and derived orbital parameters (with $1\sigma$ uncertainties) for the LP 890-9 planetary system with three different planet configurations.}
\label{tab:orbital_params}
\begin{tabular}{lccc}
\hline
Parameter & bc Solution & bdc Solution & bcd Solution \\
\hline
$M_b/M_\star$ & $(7.3 \pm 0.7) \times 10^{-5}$ & $(5.6^{+0.9}_{-0.9}) \times 10^{-5}$ & $(5.8^{+0.8}_{-0.8}) \times 10^{-5}$ \\
$M_c/M_\star$ & $(5.1^{+0.9}_{-0.9}) \times 10^{-5}$ & $(5.8^{+0.9}_{-0.9}) \times 10^{-5}$ & $(5.5^{+0.9}_{-0.9}) \times 10^{-5}$ \\
$M_d/M_\star$ & -- & $(0.62^{+0.23}_{-0.25}) \times 10^{-5}$ & $(0.38^{+0.25}_{-0.13}) \times 10^{-5}$ \\
\hline
$t_{0,b}$ (MJD) & $0.133490^{+0.000009}_{-0.000009}$ & $0.133527^{+0.000010}_{-0.000009}$ & $0.133516^{+0.000008}_{-0.000008}$ \\
$t_{0,c}$ (MJD) & $7.720317^{+0.000009}_{-0.000010}$ & $7.720263^{+0.000009}_{-0.000009}$ & $7.720280^{+0.000010}_{-0.000009}$ \\
$t_{0,d}$ (MJD) & -- & $6.24^{+0.81}_{-0.64}$ & $6.4^{+3.4}_{-1.9}$ \\
\hline
$P_b$ (days) & $2.7299058 \pm 0.0000012$ & $2.7299005 \pm 0.0000015$ & $2.7299017^{+0.0000012}_{-0.0000011}$ \\
$P_c$ (days) & $8.4574060^{+0.0000010}_{-0.0000011}$ & $8.4574048^{+0.0000012}_{-0.0000011}$ & $8.4574043 \pm 0.0000012$ \\
$P_d$ (days) & -- & $4.416^{+0.011}_{-0.008}$ & $18.53 \pm 0.06$ \\
\hline
$h_b$ & $-0.000^{+0.102}_{-0.102}$ & $-0.016^{+0.083}_{-0.076}$ & $-0.013^{+0.108}_{-0.092}$ \\
$h_c$ & $0.003^{+0.101}_{-0.103}$ & $0.002^{+0.099}_{-0.104}$ & $0.011^{+0.092}_{-0.093}$ \\
$h_d$ & -- & $-0.008^{+0.039}_{-0.041}$ & $0.004^{+0.095}_{-0.105}$ \\
\hline
$k_b$ & $-0.088^{+0.068}_{-0.044}$ & $0.018^{+0.073}_{-0.085}$ & $-0.010^{+0.095}_{-0.091}$ \\
$k_c$ & $0.110^{+0.029}_{-0.045}$ & $0.025^{+0.074}_{-0.091}$ & $0.060^{+0.058}_{-0.098}$ \\
$k_d$ & -- & $0.006^{+0.034}_{-0.042}$ & $-0.012^{+0.107}_{-0.094}$ \\
\hline
\hline
$M_b$ ($M_\oplus$) & $2.86 \pm 0.29$ & $2.19^{+0.34}_{-0.34}$ & $2.27^{+0.32}_{-0.32}$ \\
$M_c$ ($M_\oplus$) & $2.01^{+0.34}_{-0.33}$ & $2.28^{+0.33}_{-0.36}$ & $2.16^{+0.34}_{-0.36}$ \\
$M_d$ ($M_\oplus$) & -- & $0.24^{+0.09}_{-0.10}$ & $0.15^{+0.10}_{-0.05}$ \\
\hline
$e_b$ & $0.125^{+0.058}_{-0.054}$ & $0.095^{+0.058}_{-0.047}$ & $0.114^{+0.071}_{-0.057}$ \\
$e_c$ & $0.137^{+0.050}_{-0.043}$ & $0.110^{+0.068}_{-0.054}$ & $0.112^{+0.060}_{-0.050}$ \\
$e_d$ & -- & $0.047^{+0.029}_{-0.023}$ & $0.119^{+0.074}_{-0.059}$ \\
\hline
$\omega_b$ ($^{\circ}$) & $270^{+58}_{-58}$ & $134^{+104}_{-105}$ & $223^{+117}_{-109}$ \\
$\omega_c$ ($^{\circ}$) & $89^{+47}_{-45}$ & $86^{+104}_{-101}$ & $77^{+81}_{-78}$ \\
$\omega_d$ ($^{\circ}$) & -- & $152^{+105}_{-109}$ & $288^{+117}_{-116}$ \\
\hline
$T_{\mathrm{eq},d}$ (K) & -- & 335 & 208 \\
AMD-stable (\%) & -- & 65.9 & 84.2 \\
AMD-unstable (\%) & -- & 34.1 & 15.8 \\
\hline
$\chi^2/N_{data}$ & 13.8 & 3.9 & 6.0 \\
$\Delta$AIC & 188.4 & 0.0 & 40.7 \\
$\Delta$BIC & 183.4 & 0.0 & 40.7 \\
\hline
\end{tabular}

\begin{minipage}{0.95\linewidth}
\footnotesize
\textit{Notes.} Transit epochs are reported as MJD, where $\mathrm{MJD}=\mathrm{BJD}-2,460,900$.
Planet masses were computed assuming $M_\star = 0.118 \pm 0.002\,M_\odot$. For all planet-mass estimates, the uncertainty is dominated by the fitted planet-to-star mass ratio.
The argument of periapsis is computed as $\omega = \mathrm{atan2}(k,h)$; when $e$ is consistent with zero, $\omega$ becomes poorly constrained, so the large formal uncertainties on $\omega$ should be interpreted with caution. The equilibrium temperature calculation assumes zero albedo and full heat redistribution. The $\chi^2/N_{data}$, $\Delta$AIC, and $\Delta$BIC values are for the best-fit (MAP) parameter values.
\end{minipage}
\end{table*}

\subsection{Transit-Time Dynamical Fitting with \texttt{TTVFast}} 
\label{sec:analysis:ttvfast} 

As an independent check on the analytic \texttt{ttv2fast2furious} fits, we
also modeled the \texttt{Tswift} transit times using direct $N$-body integrations with \texttt{TTVFast} \citep{Deck2014}. We considered a coplanar three-planet system with planets ordered as b, d, then c, and fixed the inclinations to $i=89.5^\circ$ and the longitudes of ascending node to $\Omega=0^\circ$. The integrations were initialized five days before the first observed transit and were run through ten days after the final observed transit with a timestep of 0.02~days. The phases of planets b and c were initialized from their reference linear ephemerides, while the phase of planet d was allowed to vary. The fitted parameter vector contained 11 parameters: the mass ratios ($M/M_\star$) and the eccentricity-vector components ($e\cos\omega$ and $e\sin\omega$) for all three planets, and the orbital period and mean anomaly of planet d. We did not fit separate transit epochs or mean periods for planets b and c as sampled parameters.

The osculating periods of planets b and c at the integration start time were fixed to the periods from the joint ground-based and JWST linear ephemeris. Their mean anomalies at that time were calculated by determining the orbital phase corresponding to transit for the adopted eccentricity and argument of periastron, then propagating that phase backward from the reference transit time to $T_{\rm start}$ using the fixed period. Thus, the initial orbital elements were anchored to the reference ephemerides rather than fitted independently. Any remaining differences in the absolute transit times or mean periods were absorbed by the per-planet linear timing correction in the likelihood calculation discussed below.

For each likelihood evaluation, we compared the modeled and observed transit times, $t$, for each planet, ${p\in\{b,c\}}$:
\[
r_{p,i}(\boldsymbol{\theta}) =
    \left[t^{\rm obs}_{p,i} - 
    t^{\rm mod}_{p,i}(\boldsymbol{\theta}) \right],
\]
where \(i\) indexes the observed transit epochs and $\theta$ denotes the vector of model parameters used by \texttt{TTVFast}.
To remove sensitivity to the absolute linear ephemerides of the observed planets, at every likelihood evaluation we analytically fitted and removed a weighted linear trend in epoch, $y_{p,i} = a_p+b_pE_{p,i}$, where $E$ is the integer transit number. The fitted intercept \(a_p\) absorbs differences in the reference transit time, while the fitted slope \(b_p\) absorbs differences in the mean orbital period. The data-only likelihood then becomes:
\[
\ln\mathcal{L}_{\rm TTV}(\boldsymbol{\theta})
= -\frac{1}{2}\sum_{p\in\{b,c\}}
\min_{a_p,b_p}\sum_i\left[
\frac{r_{p,i}(\boldsymbol{\theta})-y_{p,i}
}{\sigma_{p,i}}\right]^2,
\]
where \(\sigma_{p,i}\) are the prayer-bead timing uncertainties (\autoref{sec:analysis:tswift}). 
Consequently, the likelihood constrains the nonlinear TTV structure rather than the absolute \(T_{0,p}\) and \(P_p\) values.

We sampled the posterior distribution with \texttt{emcee} \citep{emcee}, adopting physically motivated priors on the planet masses and eccentricities. For planets b and c, we used Gaussian mass priors centered on the values expected for Earth-like bulk densities, with widths broad enough to span a range of rocky compositions ($M_b=2.55 \pm 0.55 M_{\oplus}$, $M_c=2.75 \pm 0.60 M_{\oplus}$). We placed Gaussian priors centered on zero on each eccentricity-vector component, corresponding to low-eccentricity orbits expected for short-period, tidally damped planets ($\pm0.02$, $\pm0.02$, and $\pm0.06$ for planets b, c, and d). 
The independent \texttt{TTVFast} analysis presented here only considers the {\em bdc Solution}; therefore, the period of LP~890-9d was assigned a uniform prior of 4.2 -- 4.6~days.
Its orbital phase was assigned a uniform prior over
a full $360^\circ$ range. We initialized 128 walkers near the best flat-prior solution, discarded 5000 burn-in steps, and ran 100,000 production steps. We adopted the maximum-a-posteriori sample as the representative best-fit solution (\Cref{tab:map_params}), rather than the marginal posterior median, because the posterior contains correlated mass-phase degeneracies for which the component-wise median can lie away from the joint posterior ridge.

\subsection{Transit-Time Dynamical Fitting with \texttt{NbodyGradient}}
\label{sec:analysis:nbody} 

As a further check on the $bdc$ transit-timing solution, we modeled the Tswift times of transit with the gradient-aware N-body transit-timing model, \texttt{NbodyGradient.jl} \citep{Agol2021b}.  We optimized the fit using a Levenburg-Marquardt solver \citep{Gavin2019}, as implemented in \texttt{LsqFit.jl} \citep{White2026}.  As with \texttt{TTVFast}, each planet is modeled with a planet-star mass-ratio and four coplanar osculating orbital elements at the initial time $T_\mathrm{start}$ for a total of 15 free parameters. The integration time step was 0.1 days, and all four bodies were modeled using the \texttt{AHL21} algorithm \citep{Agol2021} based on \citet{Hernandez2015,Goncalves2014,Wisdom2015,Hamers2016}. The parameters were initialized with the best-fit \texttt{TTVFast} values (i.e., left-most column) in \Cref{tab:map_params}, and all parameters were allowed to vary in the fit, with Gaussian priors placed on the masses of planets $b$ and $c$ using the values listed in \Cref{tab:radii_masses}.

The model converged quickly to a chi-square of 17.5 (not including the prior terms), which is a lower chi-square than the other two models. The best-fit \texttt{Tswift}/\texttt{NbodyGradient.jl} results in \Cref{tab:map_params} are generally similar to the \texttt{Tswift}/\texttt{TTVFast} results, although the fit finds a slightly higher mass-ratio for planet $d$: $9.3\times10^{-6}$ vs.\ $6.4\times10^{-6}$, which may be due to the eccentricity-mass degeneracy which can occur when two planets are close to a 2:1 period ratio \citep{Lithwick2012,Agol2025}. 
The resulting best-fit model is shown in \Cref{fig:ttvs_nbody}.

\subsection{Orbital Stability}
\label{sec:analysis:stability} 

We evaluated the long-term orbital stability of each posterior sample using its angular momentum deficit \citep[AMD,][]{Laskar2000,Laskar2017}. AMD measures whether eccentricity growth over secular timescales can lead to orbit crossing, close encounters, or collisions between neighboring planets. Each sample was classified as either AMD-stable or AMD-unstable. AMD-stable systems cannot undergo secularly driven orbit crossing, although this criterion does not account for the stabilizing or destabilizing effects of mean-motion resonances or other chaotic short-period dynamics. AMD-unstable systems may experience orbit crossing, but are not necessarily short-lived. Systems classified as AMD-unstable therefore require a more detailed dynamical assessment, such as direct $N$-body integrations. Additionally, a system may be classified as weak AMD-stable, in which eccentricity excitation of the inner planet can lead to a collision with the star; however, we find no samples matching this classification and neglect it in further discussion.

Below we report the percentage of the 3-body systems sampled in our posteriors that are classified as either AMD-stable or AMD-unstable. We do not evaluate the 2-body solutions for stability, but note that adding a planet to a system always moves it towards the AMD-unstable regime, and so 2-body solutions are generally more AMD-stable than 3-body solutions.


\begin{table*}
    \centering
    \caption{Best-fit (i.e., MAP) values for the {\em bdc Solution} from three transit-time dynamical fits, using the \texttt{Tswift} and {\eureka} reductions as indicated. Note that the $h$ and $k$ signs reported by the \texttt{NbodyGradient} dynamical fit are reversed relative to those from the other two TTV pipelines, owing to a different line-of-sight and transit-geometry convention.}
    \begin{tabular}{lccc}
\hline
Reduction   & \texttt{Tswift}    & {\eureka} & \texttt{Tswift} \\
TTV Pipeline & TTVFast & TTV2Fast2Furious & NbodyGradient \\
\hline
$M_b/M_\star$ & $6.7 \times 10^{-5}$ & $5.5 \times 10^{-5}$ & $6.2 \times 10^{-5}$ \\
$M_c/M_\star$ & $7.1 \times 10^{-5}$ & $5.5 \times 10^{-5}$ & $7.1 \times 10^{-5}$ \\
$M_d/M_\star$ & $0.64 \times 10^{-5}$ & $0.81 \times 10^{-5}$ & $0.93 \times 10^{-5}$ \\
\hline
$t_{0,b}$ (MJD $-60900$) & -- & $0.133533$ & $0.133473$ \\
$t_{0,c}$ (MJD $-60900$) & -- & $7.720264$ & $7.720509$ \\
$t_{0,d}$ (MJD $-60900$) & $5.93$ & $5.55$ & $6.52$ \\
\hline
$P_b$ (days) & -- & $2.7299008$ & $2.7298134$ \\
$P_c$ (days) & -- & $8.4574058$ & $8.4578108$ \\
$P_d$ (days) & $4.373$ & $4.413$ & $4.366$ \\
\hline
$h_b$ & $-0.013$ & $-0.113$ & $0.014$ \\
$h_c$ & $-0.015$ & $-0.085$ & $0.051$ \\
$h_d$ & $-0.007$ & $-0.039$ & $0.022$ \\
\hline
$k_b$ & $0.002$ & $0.035$ & $0.001$ \\
$k_c$ & $-0.008$ & $0.085$ & $0.001$ \\
$k_d$ & $0.001$ & $0.047$ & $-0.005$ \\
\hline
$M_b$ ($M_\oplus$) & $2.63$ & $2.17$ & $2.44$ \\
$M_c$ ($M_\oplus$) & $2.79$ & $2.18$ & $2.78$ \\
$M_d$ ($M_\oplus$) & $0.25$ & $0.32$ & $0.36$ \\
\hline
$\rho_b$ (g cm$^{-3}$) & $5.68$ & $4.68$ & $5.27$ \\
$\rho_c$ (g cm$^{-3}$) & $5.59$ & $4.35$ & $5.57$ \\
\hline
$\chi^2$ & 21.4 & 78.4 & 17.5 \\
$\chi^2/N_{data}$ & 1.1 & 3.9 & 0.9 \\
\hline
\end{tabular}

    \label{tab:map_params}
\end{table*}

\section{Independent Constraints on LP 890-9d}

\subsection{JWST Search for Planet d} 
\label{sec:analysis:jwst}

Our first goal is to quantify the probability that the existing {\jwst} observing windows would have captured a transit of LP~890-9d, thereby constraining which orbital periods remain viable after a non-detection. This analysis assumes that LP~890-9d is transiting. For each of the 26 candidate orbital periods, we quantified the orbital-phase coverage using the start and end times of the archived {\jwst} LP~890-9 observations. We discretized the orbital phase space into 1000 bins, identified the bins that overlapped with at least one visit, and computed both a binary phase-coverage mask and a scalar coverage fraction.

The phase-coverage map in \autoref{fig:phaseCoverage} shows that the {\jwst} observations sample a large fraction of orbital phase for the short-period LP~890-9d candidates. For periods of $P_d \sim 4-7$~days, the coverage fractions span $\sim 0.5-0.8$, with the highest coverage occurring at the shortest orbital periods. In practical terms, this implies that if LP~890-9d lies between planets b and c and is transiting, there is a $>50\%$ probability that at least one transit would have fallen within the existing {\jwst} observing windows. By contrast, the phase coverage decreases steadily toward longer periods, from $\sim 0.4$ at $P_d \sim 10$~days to $<0.2$ at $P_d \gtrsim 23$~days, where large gaps in orbital phase remain unobserved. Non-detections at these longer periods are therefore much less constraining. Furthermore, the geometric transit probability of LP~890-9d decreases with increasing orbital period. Overall, \autoref{fig:phaseCoverage} shows that short-period transiting solutions are more readily testable with the existing and upcoming {\jwst} observations. For candidate periods of $P_d \sim 4$--$7$~days,  upcoming {\jwst} observations of LP~890-9c have a reasonable chance of serendipitously capturing a transit of LP~890-9d. By contrast, for $P_d \gtrsim 10$~days, the phase coverage and transit probability are substantially lower, so detecting or ruling out transits of LP~890-9d would likely require a dedicated search program.

Our second goal in this subsection is to place an upper limit on the planet radius that could be detected in a single {\jwst} observation. To do this, we searched for transits of LP~890-9d in the out-of-transit baseline of observation 51 (i.e., LP~890-9c visit 1), corresponding to the first transit of LP~890-9c. We fixed $P_d = 4.4075$~days, $i_d = 89.3^\circ$, and $a_d/R_\star = 36.2$, and placed a uniform prior of width 0.04~days on $t_{0,d}$. We find a $3\sigma$ upper limit of $(R_d/R_\star)^2 < 91$~ppm, corresponding to $R_d < 0.16\,R_\oplus$. This radius is $\sim 40\%$ smaller than the Moon and substantially smaller than the radii predicted in \autoref{sec:results}. Thus, a transiting planet with the radii predicted by our TTV solutions would have produced a visually and statistically detectable event in a single JWST visit, had the transit occurred during the observed baseline.

\subsection{TESS Search for Planet d} 
\label{sec:analysis:tess}

To search for transit signals of planet d in the TESS light curves, we analyzed detrended photometry from sectors 4, 5, 31, and 32, spanning approximately 16 weeks of observations. We first normalized each sector by the mean flux, masked outliers using a 5$\sigma$ clipping threshold with a 14.4-minute boxcar window to preserve transit features, and then concatenated the sectors in chronological order.  We masked the known transit windows of planets b and c, including padding to account for uncertainties in the transit duration. 

We phase-folded the light curve for the 26 candidate periods identified in \autoref{sec:analysis:degeneracy} and binned the 2-minute cadence data to a time resolution of one-fifth the predicted transit duration, estimated using a scaling relation from planet c ($t_{dur,2} = t_{dur,1}(P_2 / P_1)^{1/3}$).  For all candidate periods, we used a fixed ephemeris of 2460905.25 BJD. Since the transit times were unknown, the choice in ephemeris was arbitrary and did not impact our search.  For each candidate period, we identified the highest SNR dip in the binned flux by scanning through the phase-folded light curve.

Our analysis found no significant evidence for transits of planet d in the TESS light curves, and the data are not sufficiently sensitive to rule them out. All of the candidate detections had SNRs below 6$\sigma$.  When binned to the estimated transit durations, the scatter in the phase-folded light curves ranged from 700 -- 1600~ppm, which corresponds to a range of planet sizes from 0.44 -- 0.68~$R_{\oplus}$.  Thus, the available LP 890-9 TESS photometry is only marginally sensitive to detecting sub-Earth-sized planets (such as planet d).

\subsection{Independent Combined Fit} 
\label{sec:analysis:combined} 

To investigate whether the candidate LP~890-9d is independently supported by the available photometric and RV observations, we performed a series of joint fits using \texttt{PyORBIT} (v11.0.1) (see \citet{malavoltaetal16, Malavolta2018}). The fits combined the TESS photometry, ground-based transit observations from SPECULOOS and TRAPPIST-South, the {\jwst} light curves of LP~890-9b and LP 890-9c, and InfraRed Doppler (IRD) H- and YJ-band RV measurements presented by \citet{Delrez2022}.

The motivation for including the ground-based photometry was to improve the constraints on the ephemerides of the known planets and reduce degeneracies within the multi-planet fits. While several additional transit observations of LP~890-9b were publicly available, we identified no comparable ground-based transit coverage for LP~890-9c. 
For the photometric data, we used the \texttt{batman\_transit} model with quadratic limb darkening, while the RVs were modeled using the \texttt{radial\_velocities} model. The likelihood assumes independent Gaussian measurement errors, with fitted jitter terms added in quadrature to the formal uncertainties. The orbital architecture included the two confirmed planets, LP~890-9b and LP~890-9c, together with a candidate third planet, LP~890-9d. We assumed circular orbits for the known planets, in keeping with the final solution of \citet{Delrez2022}, while planet d was allowed to vary on a Keplerian orbit using the \citet{Eastman2013} parametrization, with its transit times following a linear ephemeris defined by its period and mean longitude at the reference epoch. Planet d's orbital inclination was restricted to 85$\degree$-95$\degree$, so that it was assumed to transit. Planet d was modeled as non-interacting and was included in the TESS light curve and RV models. Where required, we fitted independent photometric normalization factors, RV offsets, and jitter terms. For the {\jwst} data, we used the transit TTV implementation of \texttt{batman} \citep{Kreidberg2015} within \texttt{PyORBIT}. We fitted an independent transit time for each {\jwst} visit of LP~890-9c and adopted a linear ephemeris for LP~890-9b. Because planet d is non-interacting, the treatment of the transit times of LP~890-9b does not affect the comparison between the two- and three-planet models.

We adopted Gaussian priors on the stellar density ($37.23 \pm 3.3~\rho_\odot$), together with Gaussian priors on the derived stellar mass ($0.12 \pm 0.005$~M$_\odot$) and radius ($0.15 \pm 0.007$~R$_\odot$), based on the stellar parameters reported by \citet{Delrez2022}. The uncertainties on these parameters, as well as those on the quadratic limb-darkening coefficients, were broadened so as not to bias the results through restrictive prior constraints. The photometric normalization factors were given uniform bounds of 0.8--1.2. We estimated parameters using \texttt{emcee} \citep{emcee}, adopting 50,000 steps with 1,000 steps discarded as burn-in and a walker population four times the number of free parameters.

We explored three representative candidate orbital periods for LP~890-9d motivated by the TTV analysis (see \autoref{sec:analysis:degeneracy}), corresponding to 4.4, 5.8, and 18.4 days. We adopted uniform period priors spanning 4.2--4.6, 5.6--6.0, and 17.5--20.5 days, respectively, allowing the sampler to explore the surrounding parameter space rather than enforcing a specific candidate solution. As a control test, we additionally performed a fit including only the two confirmed planets, LP~890-9b and LP~890-9c.

The two-planet model was preferred over all three-planet configurations. While adding planet d improved the maximum likelihood by $\Delta\ln\mathcal{L}=$10--16, arising entirely from the TESS and RV data, the two-planet model has a BIC that is 54--67 lower than those of the three-planet solutions. This constitutes strong evidence against the inclusion of LP~890-9d according to commonly adopted BIC interpretation scales \citep[e.g.][]{KassRaftery1995,Trotta_2008}. The AIC, whose penalty for additional parameters does not scale with the number of data points, mildly favors the three-planet models ($\Delta$AIC\,$\sim$\,-6 to -19). The recovered transit times of LP 890-9c were identical within their uncertainties in all of our \texttt{PyORBIT} fits, as expected, since planet d does not enter the {\jwst} light curve models. Because the planet is modeled as non-interacting and the transit times of LP~890-9c are fitted independently for each visit, the {\jwst} data cannot discriminate between the two- and three-planet configurations in this framework. Although the 4.4-day fit converged to $P_{\rm d} = 4.32$\,d, the resulting solution is not physically plausible. It requires a mass of $\sim 13\,\rm{M}_\oplus$ and an eccentricity of $\sim 0.73$, which would place planet d on an orbit crossing those of both known planets. It is also inconsistent with the preferred dynamical solution in \autoref{sec:3body}. We therefore attribute this signal to noise in the TESS and RV data rather than to LP~890-9d. We also found that the inferred signal associated with LP~890-9d was sensitive to the initialization of the fit. In some runs the resulting transit-like signal was not uniquely separated from the transit signals of the known planets, with different combinations of transit signals producing comparable likelihoods, further indicating that the data do not uniquely identify a third-planet signal.

In summary, we find no evidence that LP~890-9d transits in the TESS data, and its signal cannot be identified in the IRD RVs, consistent with the conclusions of \citet{Delrez2022}. This non-detection is expected given the dynamical results of \autoref{sec:3body}. A planet of $\sim 0.2\,\rm{M}_\oplus$ at $P \sim 4.4$\,d would induce an RV semi-amplitude of only $\sim 0.4$\,m/s, well below the IRD measurement uncertainties ($\sim8-12$\,m/s) and fitted jitter ($\sim 10-40$\,m/s). If rocky and transiting, such a planet would produce a transit depth that is below the sensitivity of the available TESS photometry (\autoref{sec:analysis:tess}). This independent analysis therefore provides a consistent observational basis for comparison with the dynamical models presented in \autoref{sec:3body}, but does not by itself uniquely identify the presence of planet d or the preferred orbital configuration. We recommend continued photometric and spectroscopic monitoring of this system to further constrain the masses and architecture of the LP~890-9 system.

\section{Discussion} 
\label{sec:discussion}

The transit-time dynamical fits presented above provide evidence for an additional perturbing planet in the LP~890-9 system, but they do not yet constitute a unique or definitive orbital solution.
The measured TTVs appear to be most robust for planet c. The two-planet {\em bc Solution} cannot reproduce either the observed super-period or the amplitude of the planet c TTVs, whereas both three-planet configurations provide substantially better agreement. The information criteria favor the {\em bdc Solution}, with $\Delta{\rm AIC}=\Delta{\rm BIC}=40.7$ for the {\em bcd Solution}. This difference primarily stems from the {\em bcd Solution} failing to reproduce the measured TTVs of planet b. Thus, if the measured planet b TTVs are astrophysical, then the orbital ordering should have planet d located between planets b and c (i.e., the {\em bdc Solution}). However, a range of orbital periods, from 4.0 -- 6.9 days, remain plausible and the current preference for 4.40 days could change with additional data.

Although short-period solutions have relatively high phase coverage in the existing {\jwst} observations, the absence of a detected transit does not rule out this architecture if LP~890-9d is modestly non-coplanar. For the favored $P_d\simeq4.40$~day solution, a mutual inclination of only $\sim1^\circ$ relative to the transiting planets would be sufficient to move LP~890-9d outside the transit chord, making a non-transiting interior perturber dynamically plausible.

In \Cref{tab:map_params}, we compare the best-fit, or maximum-a-posteriori (MAP), values for the {\em bdc Solution} from three different transit-time dynamical fitting pipelines. The \texttt{TTVFast} and NbodyGradient pipelines achieve similarly good fits ($\chi^2/N_{data} \sim 1$) and favor an orbital period of $\sim 4.37$~days. For all three pipelines, the inferred mass of LP~890-9d is broadly comparable, ranging from 0.25 -- 0.36~$M_\oplus$.

Additional transit timing measurements are needed to determine whether the planet b TTVs are real and to break the remaining degeneracy in the orbit of LP~890-9d. {\jwst} transit observations in August -- October 2026 will extend the timing baseline of planet c. However, because these observations do not include transits of planet b, they will not directly determine whether the planet b TTVs are astrophysical; confirming those timings remains necessary to establish whether an interior perturber in the {\em bdc} configuration is required.

A majority of posterior samples in both three-planet configurations are AMD-stable, indicating that neither solution is immediately disfavored by secular stability. The posterior for the {\em bcd Solution} has a larger AMD-stable fraction, but this does not overcome its poorer fit to the planet-b TTVs.

\section{Conclusion}
\label{sec:conclusion}

In this work, we analyzed 20 {\jwst} transits of LP~890-9b and LP~890-9c and found evidence for TTVs that favor the presence of an additional planet candidate, LP~890-9d. Our main findings are as follows:

\begin{itemize}
\item{\textbf{The two-planet solution is inadequate.}}
    The observed {\jwst} TTVs have peak-to-peak amplitudes that are substantially larger than the amplitudes predicted for the known two-planet system. The fitted {\em bc Solution} provides the poorest fit to the data and cannot reproduce the observed super-period.

\item{\textbf{The preferred configuration places LP~890-9d between planets b and c.}}
    Both three-planet configurations improve the fit to the LP~890-9c TTVs; however, only the {\em bdc Solution} can reproduce the measured TTVs of LP~890-9b. While the orbital parameters of LP~890-9d are poorly constrained, both information-criterion values strongly favor the {\em bdc Solution}.

\item{\textbf{LP~890-9d is a low-mass, sub-Earth-size planet candidate with a degenerate orbital period.}}
    The orbital period of LP~890-9d is likely associated with one of the 12 candidate periods from 4.0--6.9 days that were identified from the TTV super-period analysis. While the phase coverage ranges from 50--80\% over this orbital period range, no transits of LP~890-9d were seen in the existing {\jwst} observations. Follow-up observations with {\jwst} or a large ground-based telescope would have sufficient precision to search for transits of LP~890-9d.
\end{itemize}

This work demonstrates that {\jwst} observations of compact multi-planet systems can provide valuable science beyond the primary observing goal. In particular, high-precision transit observations obtained for atmospheric characterization can also enable dynamical studies, including TTV searches, mass constraints, and searches for additional planets. These synergies should motivate future {\jwst} programs designed to serve multiple science cases simultaneously, combining atmospheric characterization with system-level constraints on planetary dynamics and architectures.

\begin{acknowledgments}

We would like to thank our anonymous reviewer for their constructive comments and suggestions. 
This work is based on observations made with the NASA/ESA/CSA James Webb Space Telescope. 
The data were obtained from the Mikulski Archive for Space Telescopes at the Space Telescope Science Institute, which is operated by the Association of Universities for Research in Astronomy, Inc., under NASA contract NAS 5-03127 for JWST. These observations are associated with program \#7073. 
L.P. acknowledges funding from the Royal Society Career Development Fellowship, grant number CDF\textbackslash
R1\textbackslash251054.
The specific observations analyzed can be accessed via \dataset[doi:10.17909/3f8b-zd40]{https://doi.org/10.17909/3f8b-zd40}.
We made use of OpenAI’s ChatGPT to assist with generating Python code, creating tables and figures, and improving clarity in parts of this manuscript.

\textbf{AI Usage Disclosure.} The authors used large language models (LLMs), primarily OpenAI's ChatGPT and Anthropic's Claude, to assist with various aspects of this work, including code development and debugging, implementation and refinement of the TTVFast and TTV2Fast2Furious analyses, preparation of figures and tables, and manuscript copy editing. All AI-assisted analyses, calculations, and scientific interpretations were reviewed and validated by the authors, who take full responsibility for the accuracy and integrity of the results and final manuscript.

\end{acknowledgments}

\facilities{JWST (NIRSpec)}

\software{
\texttt{Astraeus} \citep{astraeus};
\texttt{AstroPy} \citep{astropy2013};
\texttt{batman} \citep{Kreidberg2015};
\texttt{CRDS} \citep{crds};
\texttt{emcee} \citep{emcee};
\texttt{ExoTiC-LD} \citep{Grant2024};
\eureka \citep{Bell2022};
\texttt{fleck} \citep{fleck};
\texttt{h5py} \citep{h5py};
\texttt{jwst} \citep{jwst};
\texttt{Matplotlib} \citep{Hunter2007}; 
\texttt{NumPy} \citep{numpy};
\texttt{SciPy} \citep{scipy};
\texttt{Xarray} \citep{hoyer2017xarray}
}.

\appendix

\begin{table}
\centering
\caption{Best-fit resonance solutions per family from the phase-space scan in \autoref{sec:analysis:degeneracy}. Here $j$ and $k$ are the resonance integers defining the $j:k$ mean-motion resonance between LP~890-9c and LP~890-9d.}
\label{tab:pd_resonance}
\begin{tabular}{c c c c c}
\hline\hline
$P_d$ (days) & $P_d/P_c$ & $j$ & $k$ & $j-k$ \\
\hline
4.065611 & 0.480716 & 2 & 1 & 1 \\
4.145531 & 0.490166 & 4 & 2 & 2 \\
4.315363 & 0.510247 & 4 & 2 & 2 \\
4.405274 & 0.520878 & 2 & 1 & 1 \\
4.994320 & 0.590526 & 5 & 2 & 3 \\
5.157122 & 0.609776 & 5 & 2 & 3 \\
5.491605 & 0.649325 & 3 & 1 & 2 \\
5.793158 & 0.684980 & 3 & 1 & 2 \\
6.218292 & 0.735248 & 4 & 1 & 3 \\
6.472855 & 0.765347 & 4 & 1 & 3 \\
6.658967 & 0.787353 & 5 & 1 & 4 \\
6.876159 & 0.813034 & 5 & 1 & 4 \\
10.363814 & 1.225412 & 5 & 1 & 4 \\
10.788208 & 1.275593 & 5 & 1 & 4 \\
10.982830 & 1.298605 & 4 & 1 & 3 \\
11.586306 & 1.369959 & 4 & 1 & 3 \\
12.197182 & 1.442189 & 3 & 1 & 2 \\
13.216172 & 1.562674 & 3 & 1 & 2 \\
13.728627 & 1.623266 & 5 & 2 & 3 \\
14.482695 & 1.712427 & 5 & 2 & 3 \\
15.658937 & 1.851505 & 2 & 1 & 1 \\
16.262783 & 1.922904 & 4 & 2 & 2 \\
17.621436 & 2.083550 & 4 & 2 & 2 \\
18.389564 & 2.174373 & 2 & 1 & 1 \\
23.488585 & 2.777279 & 3 & 2 & 1 \\
27.584526 & 3.261581 & 3 & 2 & 1 \\
\hline
\end{tabular}
\end{table}

\begin{figure*}[t]
    \centering
    \includegraphics[width=\textwidth]{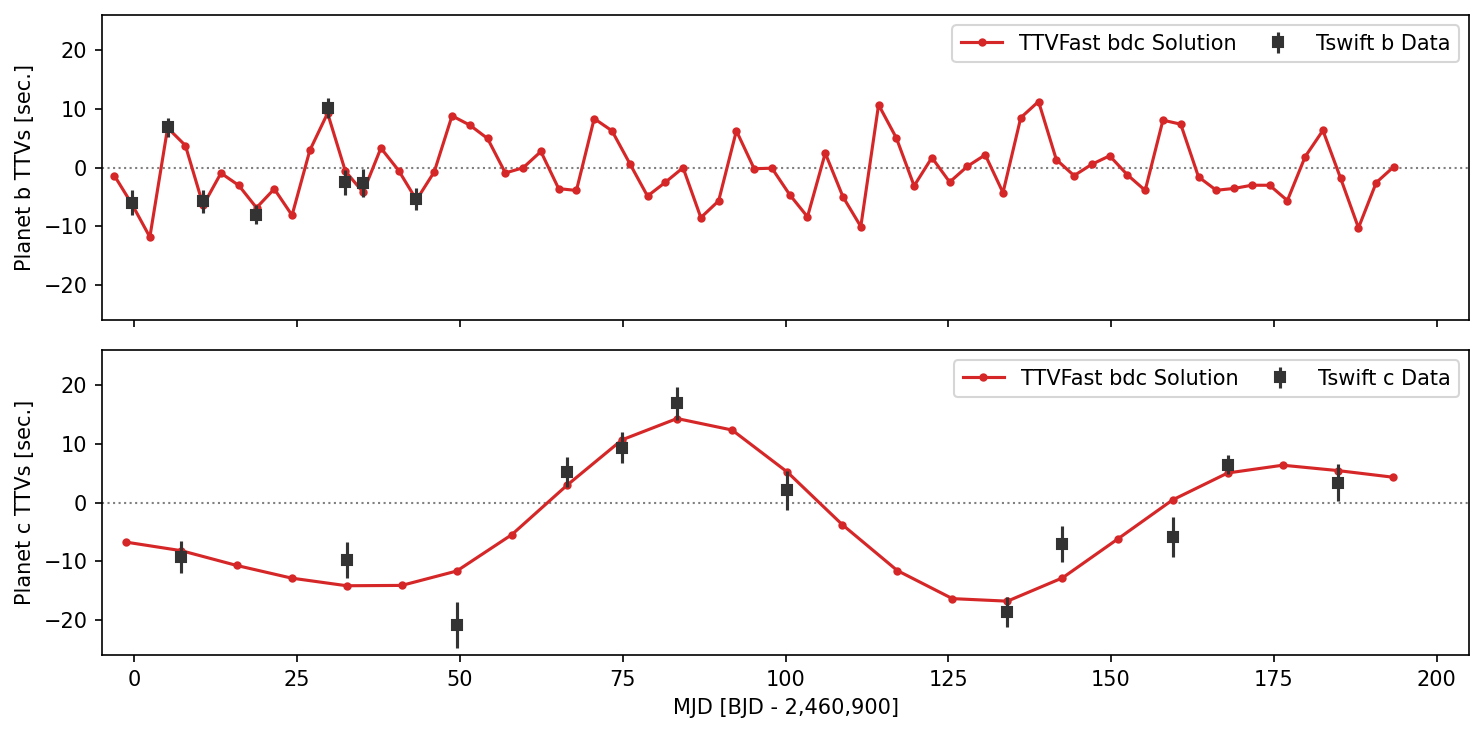}
    \caption{Measured \texttt{Tswift} TTVs for LP~890-9b (top) and LP~890-9c (bottom), shown as black squares, compared with the best-fit (MAP) model solutions, shown as colored points connected by lines, using \texttt{TTVFast}. 
    }
    \label{fig:ttvs_ttvfast}

    \vspace{\floatsep}
    
    \includegraphics[width=\textwidth]{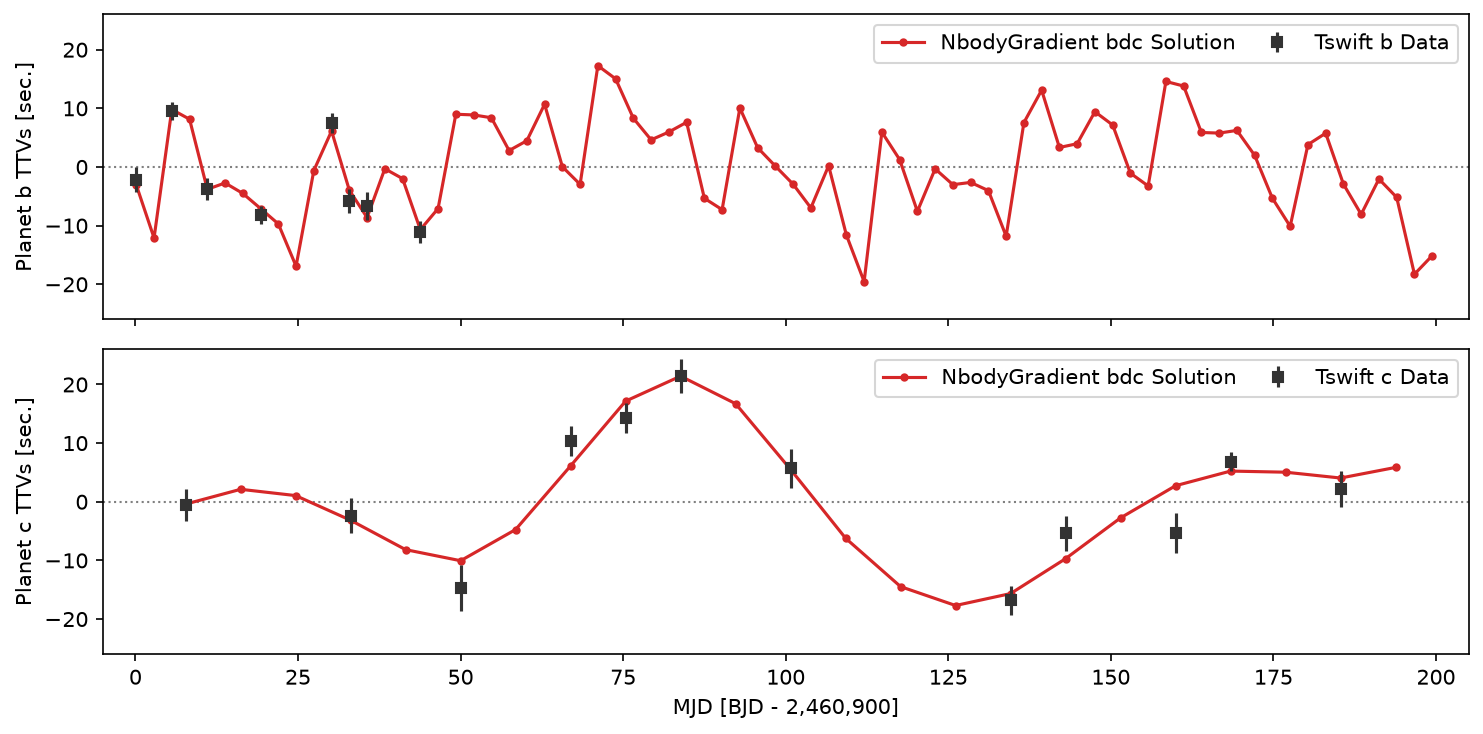}
    \caption{Measured \texttt{Tswift} TTVs for LP~890-9b (top) and LP~890-9c (bottom), shown as black squares, compared with the best-fit (MAP) model solutions, shown as colored points connected by lines, using NbodyGradient. 
    }
    \label{fig:ttvs_nbody}
\end{figure*}

\bibliography{main}

\end{document}